\documentclass[conference]{IEEEtran}
\IEEEoverridecommandlockouts

\usepackage{cite}
\usepackage{amsmath,amssymb,amsfonts}
\usepackage{algorithmic}
\usepackage{graphicx}
\usepackage{textcomp}
\usepackage{xcolor}
\def\BibTeX{{\rm B\kern-.05em{\sc i\kern-.025em b}\kern-.08em
    T\kern-.1667em\lower.7ex\hbox{E}\kern-.125emX}}

\usepackage{hyperref}
\usepackage{graphicx}
\usepackage{amssymb,amsmath,bm}
\usepackage{textcomp}
\usepackage{booktabs}
\usepackage[textfont=it,tableposition=top]{caption}
\usepackage{siunitx}
\usepackage{xspace}
\usepackage{url}
\usepackage{lipsum}  
\usepackage{tipa}
\usepackage{enumitem}
\usepackage{xkeyval}
\usepackage{calc}
\usepackage[capitalize]{cleveref}
\usepackage{booktabs}
\usepackage{multirow}
\usepackage{subcaption}
\usepackage{cite}
\usepackage{microtype}

\crefname{section}{Sec.}
\makeatletter

\newcommand*{\ie}{i.e.\@\xspace}

\makeatother

\begin{document}

\title{Autoregressive Speech Enhancement via \\ Acoustic Tokens}

\author{\IEEEauthorblockN{Luca Della Libera$^{1,2}$, Cem Subakan$^{3,1,2}$, Mirco Ravanelli$^{1,2}$}
\IEEEauthorblockA{\textit{$^1$Concordia University, 
  $^2$Mila-Quebec AI Institute,
  $^3$Université Laval}\\
luca.dellalibera@mail.concordia.ca, cem.subakan@ift.ulaval.ca, mirco.ravanelli@concordia.ca}
}

\maketitle

\begin{abstract}
In speech processing pipelines, improving the quality and intelligibility of real-world recordings is crucial. While supervised regression is the primary method for speech enhancement, audio tokenization is emerging as a promising alternative for a smooth integration with other modalities.
However, research on speech enhancement using discrete representations is still limited. Previous work has mainly focused on semantic tokens, which tend to discard key acoustic details such as speaker identity. Additionally, these studies typically employ non-autoregressive models, assuming conditional independence of outputs and overlooking the potential improvements offered by autoregressive modeling.
To address these gaps we: 1) conduct a comprehensive study of the performance of acoustic tokens for speech enhancement, including the effect of bitrate and noise strength; 2) introduce a novel transducer-based autoregressive architecture specifically designed for this task. Experiments on VoiceBank and Libri1Mix datasets show that acoustic tokens outperform semantic tokens in terms of preserving speaker identity, and that our autoregressive approach can further improve performance. Nevertheless, we observe that discrete representations still fall short compared to continuous ones, highlighting the need for further research in this area.
\end{abstract}


\section{Introduction}
Speech enhancement aims to improve the quality of a speech signal by removing the noise and reverberation commonly found in real-life recordings.
In recent years, significant progress has been made in this domain.
While unsupervised~\cite{fu2022metricganu}, semi-supervised~\cite{fu2019metricgan,fu2021metricganplus}, and reinforcement learning-based~\cite{shen2019reinforcement,zhou2023metarl} methods have been proposed to address speech enhancement, the dominating paradigm so far has been supervised regression. In this framework, \emph{continuous} speech features serve as input/output, and reconstruction losses such as L1 and L2 or perception-based metrics such as PESQ~\cite{rix2001pesq} and STOI~\cite{taal2011stoi}
are used to align the prediction with the target.
In particular, techniques working in the time domain~\cite{luo2019conv,subakan2020attention,defossez2020demucs} and time-frequency domain~~\cite{zhao2016dnn, yin2020phasen, chen2023intersubnet, welker2022speech, richter2023speech} have been the most successful.

A novel approach in speech processing is \emph{audio tokenization}, which converts audio signals into discrete tokens, allowing them to be processed similarly to text using natural language processing techniques.
This unified framework enables seamless integration between audio and text modalities, facilitating the development of multi-modal systems~\cite{geminiteam2023gemini}.
As a result, audio tokenization is increasingly being adopted in tasks like text-conditioned speech and music generation, as seen in models like AudioLM~\cite{borsos2023audiolm}, VALL-E~\cite{wang2023valle}, AudioGen~\cite{kreuk2023audiogen}, MusicGen~\cite{copet2023musicgen}, SpeechGPT~\cite{zhang2023speechgpt}, SpiRit-LM~\cite{nguyen2024spiritlminterleavedspokenwritten}, NaturalSpeech 3~\cite{ju2024naturalspeech}, and CLaM-TTS\cite{kim2024clamtts}.
Speech tokens can be broadly classified into two categories: 1) \emph{semantic}, which are typically obtained by applying k-means clustering to large pretrained self-supervised speech encoders~\cite{baevski2020wav2vec2,hsu2021hubert,chen2022wavlm} and capture linguistic details, making them suitable for tasks like speech recognition and spoken language understanding~\cite{chang2023exploring,yang2024universal}; 2) \emph{acoustic}~\cite{defossez2023encodec, kumar2023dac, du2023funcodec, yang2023hifi, ji2024languagecodec, ju2024naturalspeech, ji2024wavtokenizerefficientacousticdiscrete}, which are typically obtained by applying residual vector quantization~\cite{vandenord2017vqvae,zeghidour2021soundstream} to audio autoencoders and aim to retain all information, making them suitable for tasks like multi-speaker text-to-speech and speaker verification.
On the one hand, the k-means discretization process often leads to the loss of important acoustic details, such as speaker identity~\cite{vanniekerk2022}. In contrast, acoustic tokens preserve richer audio information but at the cost of higher bitrates, resulting in long sequences.
A promising direction is the development of \emph{hybrid} tokens~\cite{zhang2024speechtokenizer, liu2024semanticodec}, which aim to balance the strengths of both approaches.

While audio tokenization has been applied to various tasks, research on its use in speech enhancement remains limited. Most studies~\cite{wang2024selm, mousavi2024how} have focused exclusively on semantic tokens, neglecting the potential of acoustic tokens, which are more suited for preserving speaker identity. Although \cite{mousavi2024dasb} includes an evaluation of acoustic tokens, it relies on simple baselines, uses only a single dataset, and lacks analysis of the effect of noise strength. Additionally, all these works assume \emph{conditional independence} of outputs, overlooking the improvements that autoregressive modeling could offer.
In this study, we aim to bridge these gaps with the following contributions:
\begin{itemize}[leftmargin=0.35cm]
    \item We analyze the performance of acoustic tokens for speech enhancement on VoiceBank and Libri1Mix, including the effect of bitrate and noise strength. We show that acoustic tokens outperform semantic tokens in preserving speaker identity.
    \item We introduce a novel autoregressive architecture, the \textbf{Speech Enhancement Transducer}, that draws inspiration from transducers~\cite{graves2012transducer} used in speech recognition. We show that while autoregressive modeling is beneficial, it is also more prone to exposure bias~\cite{he2021exposure}, necessitating the use of techniques to mitigate this issue.
\end{itemize}

\begin{figure*}[t]
  \centering
\includegraphics[width=0.625\textwidth]{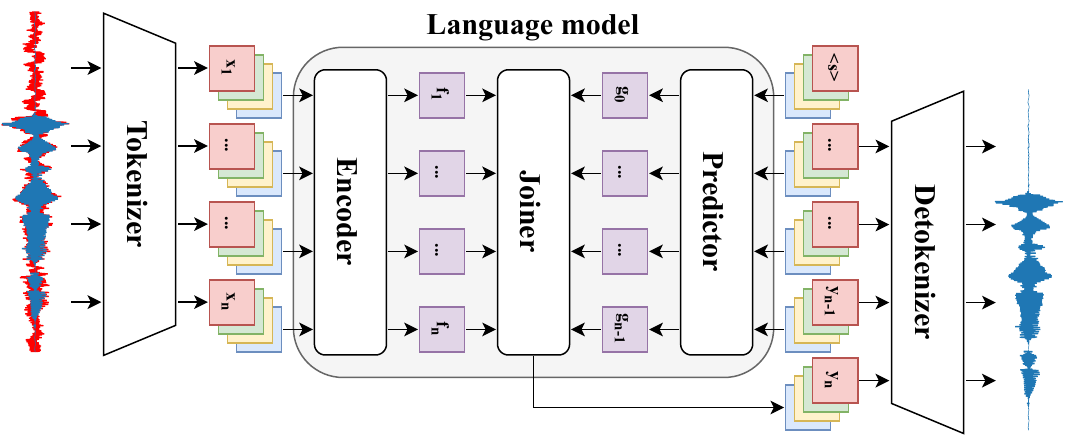}
  \vspace{-.2cm}
  \caption{\small Overview of our speech enhancement framework, consisting of a tokenizer, a language model, and a detokenizer. The language model is based on the proposed \textbf{Speech Enhancement Transducer} (SET) autoregressive architecture.}
  \label{fig:framework}
  \vspace{-0.6cm}
\end{figure*}

\section{Method}

\subsection{Framework Overview}
As shown in \cref{fig:framework}, our system consists of a tokenizer that converts a raw noisy waveform into discrete tokens, a language model that transforms the sequence of noisy tokens into a sequence of enhanced tokens, and a detokenizer that converts the enhanced speech tokens back to the time domain into an enhanced waveform.

\subsection{Tokenizer}
The tokenizer is responsible for converting raw audio into tokens.
For acoustic tokens, we use the pretrained residual vector quantizer included with the codec.
For semantic tokens, we follow the literature~\cite{chang2023exploring,yang2024universal,wang2024selm,mousavi2024how} and train our own k-means quantizer on top of the continuous representations extracted from a large pretrained self-supervised speech encoder.

\subsection{Language Model}
The language model is the core component of our system, whose goal is to convert noisy input tokens into enhanced tokens.
While autoregressive encoder-decoder transformer~\cite{vaswani2017transformer} architectures have been proposed for sequence-to-sequence tasks like text-conditioned speech generation, we argue that this approach is suboptimal for the task at hand. Indeed, speech enhancement can be better modeled as a many-to-many problem rather than a sequence-to-sequence one, where \emph{each} token in the input corresponds to an enhanced token in the output.
Thus, employing a full encoder-decoder transformer framework with cross-attention is inefficient in terms of model capacity utilization, as no alignment needs to be learned.
Based on this observation, \cite{wang2024selm,mousavi2024how,mousavi2024dasb} employ a non-autoregressive transformer encoder, trained via cross-entropy loss.
However, predictions obtained through this approach are conditionally independent, which means that the output at the current time step does not \emph{explicitly} depend on the output at the previous time steps.
Hence, we hypothesize that a more effective model $h$ could be developed by leveraging information from past outputs:
\begin{equation}
y_n = h(x_1, \dots, x_n, \, y_1, \dots, y_{n-1}),
\end{equation}
where $x_1, \dots, x_n$ is the sequence of noisy tokens, and $y_1, \dots, y_n$ the sequence of clean tokens.
Drawing inspiration from successful methods in speech recognition, we introduce the \textbf{Speech
Enhancement Transducer} (SET), a novel autoregressive speech enhancement model. It consists of three modules: an \emph{encoder}, a \emph{predictor}, and a \emph{joiner}. The encoder extracts features from the noisy input sequence, while the autoregressive predictor processes the enhanced output tokens up to the previous time step, providing features useful for predicting the enhanced token at the current time step. Then, the encoder and predictor hidden representations are combined by the joiner via element-wise sum or concatenation, as in the widely adopted transducer~\cite{graves2012transducer} architecture used for speech recognition. The model is trained end-to-end via cross-entropy loss.

\subsection{Detokenizer}
The detokenizer handles the task of converting the enhanced token sequence back to the time domain.
Essentially, it is a vocoder that operates on discrete units instead of spectrograms.
For acoustic tokens, it is not necessary to train a detokenizer, as they natively include a pretrained one, fine-tuned along with the quantizer in an end-to-end fashion.
For semantic tokens, we train our own detokenizer.

\section{Experimental Setup}

\subsection{Datasets} 
\begin{itemize}[label={},leftmargin=0pt]
    \item \textbf{VoiceBank}~\cite{valentinibotinhao2016voicebank}. The dataset includes 11,572 utterances from 28 speakers in the training set (noise at 0 dB, 5 dB, 10 dB, and 15 dB), and 872 utterances from 2 unseen speakers in the test set (noise at 2.5 dB, 7.5 dB, 12.5 dB, and 17.5 dB).
    To form a validation set, we randomly select one speaker from the training set (353 utterances).
    \item \textbf{Libri1Mix}~\cite{cosentino2020librimix}. The dataset consists of a training set (\texttt{train-100}), a validation set (\texttt{dev}), and a test set (\texttt{test}), randomly selected from the corresponding subsets of LibriSpeech~\cite{panayotov2015librispeech}. The clean utterances from the first speaker are mixed with noise from the WHAM!~\cite{wichern2019wham} corpus with SNRs uniformly distributed between 0 and 5 dB.
\end{itemize}

\subsection{Compared Methods}
\label{subsec:compared_methods}
\begin{itemize}[label={},leftmargin=0pt]
\item \textbf{Codecs}. For acoustic codecs, we consider \textbf{EnCodec}~\cite{defossez2023encodec} and \textbf{DAC}~\cite{kumar2023dac}. In particular, we use the 24 kHz monophonic variant with a bitrate of 3.0 kbps (4 codebooks) and a codebook size of 1024. The pretrained quantizer and vocoder are employed as the tokenizer and detokenizer, respectively.
For EnCodec, we also consider replacing the default vocoder with \textbf{Vocos}~\cite{siuzdak2023vocos}, which generates higher-quality audio reconstructions.
As a semantic codec, we consider \textbf{WavLM} \textbf{discrete}~\cite{chen2022wavlm}. Following \cite{wang2024selm}, as the tokenizer we use a k-means quantizer with 512 clusters trained on representations extracted from the 6-th transformer layer of the large variant of WavLM\footnote{\href{https://huggingface.co/microsoft/wavlm-large}{https://huggingface.co/microsoft/wavlm-large}}~\cite{chen2022wavlm}. For the detokenizer, we first train a dequantizer via L2 loss to map the discrete representations back to continuous ones, and then train a vocoder on top of these reconstructed continuous representations.
For the dequantizer architecture, we use a Conformer~\cite{gulati2020conformer} encoder with 6 layers, 4 attention heads, a model dimension of 256, a feed-forward layer dimension of 256, and a dropout probability of 0.1.
For the vocoder architecture, we use HiFi-GAN V1~\cite{jungil2020hifigan}.
Both quantizer, dequantizer and vocoder are trained on LibriSpeech~\cite{panayotov2015librispeech} \texttt{train-clean-100} split.\looseness-1

\item \textbf{Architectures}.
For the language model architecture, two variants are considered: \textbf{NAR} (non-autoregressive) and \textbf{AR} (autoregressive). The NAR variant uses a Conformer encoder with 6 layers, 4 attention heads, a model dimension of 256, a feed-forward layer dimension of 2048, and a dropout probability of 0.1. The input is obtained by feeding the noisy tokens into $K$ embedding layers, one for each codebook, each with $C$ entries, and summing over the codebook dimension. The cross-entropy logits are computed using $K$ linear layers, each with $C$ output channels.
The AR variant employs the proposed SET architecture. The predictor is a single causal Conformer layer with 4 attention heads, a model dimension of 256, a feed-forward layer dimension of 2048, and a dropout probability of 0.1. The predictor's input is obtained by feeding clean tokens (with the start-of-sequence token \texttt{<s>} prepended) into $K$ embedding layers, each with $C + 1$ entries, and summing over the codebook dimension. The joiner is a single linear layer with 120 channels followed by an element-wise sum. To ensure a fair comparison, the number of layers in the AR encoder is reduced from 6 to 5, to approximately match the size of the NAR model. The model is trained end-to-end using cross-entropy loss with teacher forcing, which is disabled for the final 5 epochs to mitigate exposure bias (see \cref{subsec:exposure_bias}). At inference, beam search with 5 beams is used.

\item \textbf{Continuous representations}. We consider \textbf{WavLM} \textbf{continuous}~\cite{chen2022wavlm} and \textbf{SepFormer}~\cite{subakan2020attention} as continuous representations baselines. For WavLM continuous, we turn the language model into a regressor that learns to map noisy representations from the 6-th layer of WavLM to their corresponding enhanced representations via L2 loss minimization.
For SepFormer, encoder and decoder are based on 256 convolutional filters with a kernel size of 16 a stride factor of 8. The masking network consists of 2 SepFormer blocks with a chunk size of 250. Each block includes 4 intra-transformer and 4 inter-transformer layers with 4 attention heads, a model dimension of 256, a feed-forward layer dimension of 2048, and a dropout probability of 0.1. The model is trained end-to-end via negative scale-invariant signal-to-noise ratio~\cite{le2019sdr} loss.
\end{itemize}

\begin{table*}[t]
\newcommand\TBstrut{\rule[-0.7ex]{0pt}{3.2ex}}
\centering
\caption{\small Evaluation results. The best values for each codec are highlighted in \textbf{bold}. The best overall values are \setlength{\fboxsep}{1pt}\fbox{\textbf{framed}}.}
\vspace{-0.1cm}
\label{tab:quantitative_results}
\resizebox{1.0\textwidth}{!}{%
\begin{tabular}{l|c|c|c|c|c|c|c|c|c}
\hline
\multirow{2}{*}{\textbf{Method}} & \multirow{2}{*}{\textbf{Category}} & \multicolumn{3}{c|}{\textbf{VoiceBank}} & \multicolumn{3}{c|}{\textbf{Libri1Mix}} & \multicolumn{2}{c}{\textbf{\#Params (M)}} \\ \cline{3-10}
& & \textbf{DNSMOS} ($\uparrow$) & \textbf{CosSim} ($\uparrow$) & \textbf{dWER} ($\downarrow$) & \textbf{DNSMOS} ($\uparrow$) & \textbf{CosSim} ($\uparrow$) & \textbf{dWER} ($\downarrow$) &
\textbf{Codec} & \textbf{Model} \TBstrut \\
\hline
\hline
EnCodec NAR & Acoustic & \textbf{3.19} & 0.872 & 30.21 & \textbf{3.31} & \textbf{0.908} & 48.26 & 24 & 19 \\

EnCodec AR & Acoustic & 3.18 & \textbf{0.877} & \textbf{23.97} & 3.30 & {0.904} & \textbf{45.81} & 24 & 20 \\ \hline

EnCodec NAR + Vocos & Acoustic & \textbf{3.46} & 0.916 & 22.04 & \textbf{3.68} & \textbf{0.919} & 46.87 & 50 & 19 \\

EnCodec AR + Vocos & Acoustic & 3.45 & \textbf{0.927} & \textbf{21.69} & 3.66 & 0.915 & \textbf{46.29} & 50 & 20 \\ \hline

DAC NAR & Acoustic & \textbf{3.41} & {0.896} & \textbf{31.18} & 3.38 & 0.894 & 63.22 & 74 & 19 \\

DAC AR & Acoustic & \textbf{3.41} & \textbf{0.899} & 31.69 & \textbf{3.40} & \textbf{0.899} & \textbf{60.01} & 74 & 20 \\ \hline

WavLM discrete NAR & Semantic & \textbf{3.42} & \textbf{0.837} & \textbf{22.39} & 3.65 & \textbf{0.888} & \textbf{38.65} & 127 & 17 \\

WavLM discrete AR & Semantic & {3.41} & {0.833} & {22.77} & \textbf{3.66} & \textbf{0.888} & 38.71 & 127 & 20 \\ \hline

WavLM continuous & Continuous & \setlength{\fboxsep}{1pt}\fbox{\textbf{3.52}} & 0.934 & 4.12 & \setlength{\fboxsep}{1pt}\fbox{\textbf{3.77}} & 0.959 & 14.00 & 107 & 17 \\ \hline

SepFormer & Continuous & 3.35 & \setlength{\fboxsep}{1pt}\fbox{\textbf{0.968}} & \setlength{\fboxsep}{1pt}\fbox{\textbf{3.67}} & 3.36 & \setlength{\fboxsep}{1pt}\fbox{\textbf{0.962}} & \setlength{\fboxsep}{1pt}\fbox{\textbf{10.58}} & - & 24 \\ \hline
\end{tabular}
}
\vspace{-0.25cm}
\end{table*}

\subsection{Evaluation Metrics}
To measure signal quality, we use the deep noise suppression mean opinion score (\textbf{DNSMOS})~\cite{reddy2022dnsmos}.
To measure speaker fidelity, we use the cosine similarity (\textbf{CosSim}) between x-vectors extracted from the enhanced signal and the target signal using the base variant of WavLM fine-tuned for speaker verification\footnote{\href{https://huggingface.co/microsoft/wavlm-base-sv}{https://huggingface.co/microsoft/wavlm-base-sv}}~\cite{chen2022wavlm}.
To measure intelligibility, we use the differential word error rate (\textbf{dWER})~\cite{wang2021dwer}, \ie the word error rate between the transcribed enhanced signal and the transcribed clean signal. To obtain the transcriptions, we use the small variant of Whisper\footnote{\href{https://huggingface.co/openai/whisper-small}{https://huggingface.co/openai/whisper-small}}~\cite{radford2022robust}.\looseness=-1

\subsection{Training Details}
All Conformer models were trained to convergence with mixed precision on variable-duration audio batches (up to 90 seconds) for up to 50/25 epochs on VoiceBank/Libri1Mix, using the AdamW optimizer with an initial learning rate of 5e\scalebox{0.75}[1.0]{$-$}4 and a weight decay of 1e\scalebox{0.75}[1.0]{$-$}2. Learning rate is annealead by a multiplicative factor of 0.9 if no validation loss improvement is observed, and gradient L2 norm is clipped to 5. Data augmentation includes randomly dropping frequency bands and/or chunks with a 0.75 probability. The k-means quantizer is trained until cluster centers stop changing, with larger batches (up to 200 seconds) to improve stability. The HiFi-GAN vocoder is trained according to the original recipe~\cite{jungil2020hifigan}.
Software for the experimental evaluation was implemented in Python using the SpeechBrain~\cite{ravanelli2021speechbrain, ravanelli2024speechbrain} toolkit. All the experiments were run on a CentOS Linux machine with an Intel(R) Xeon(R) Silver 4216 Cascade Lake CPU with 32 cores @ 2.10 GHz, 32 GB RAM and an NVIDIA Tesla V100 SXM2 @ 32 GB with CUDA Toolkit 11.7.\looseness=-1

\section{Results}

\subsection{Comparative Study}
\label{subsec:comparative_study}
\cref{tab:quantitative_results} shows the results of our comparative study.
We observe that acoustic tokens from EnCodec and DAC outperform semantic tokens from WavLM with respect to speaker fidelity, as indicated by the higher cosine similarity across all settings.
This aligns with expectations, as the k-means quantization tends to remove speech features like prosody and intonation.
Among acoustic codecs, DAC outperforms EnCodec by a large margin in terms of signal quality and speaker fidelity at the considered bitrate (3.0 kbps). However, the generated speech is less intelligible.
Interestingly, {EnCodec + Vocos} outperforms DAC. This highlights how training a specialized vocoder can enhance quality without fine-tuning the quantizer.
We also note that our SET architecture, on average, improves intelligibility. However, on Libri1Mix, while the enhancement effect is still evident as indicated by high DNSMOS values, intelligibility suffers, due to the higher noise levels compared to VoiceBank.
Despite achieving competitive performance, it is worth noting that methods based on discrete representations lag behind continuous ones, especially with respect to intelligibility. While the top-performing codec is competitive against the continuous variant of WavLM in terms of DNSMOS, and even superior to SepFormer, it obtains significantly worse dWER. We hypothesize that this performance gap is due to the large amount of data required to effectively train a language model on high-bitrate tokens. For instance, other studies on discrete audio representations for speech enhancement and separation utilize thousands of hours of data~\cite{wang2024selm,erdogan2023tokensplit}, whereas, in our case, we utilize no more than a hundred hours. Furthermore, while architectures like SepFormer are specifically designed for speech enhancement and separation tasks, our model is more versatile and can potentially accommodate multiple modalities.
\looseness=-1

\begin{figure}[t]
    \vspace{-0.4cm}
    \begin{center}
    \begin{subfloat}{}
    \includegraphics[width=0.20\textwidth]{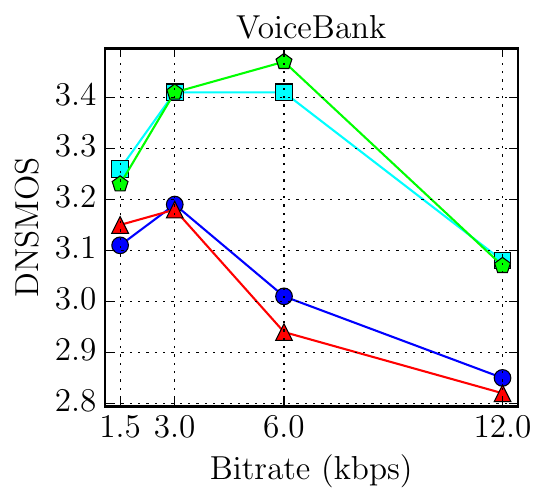}
    \end{subfloat}
    \vspace{-0.2cm}
    \begin{subfloat}{}
    \includegraphics[width=0.20\textwidth]{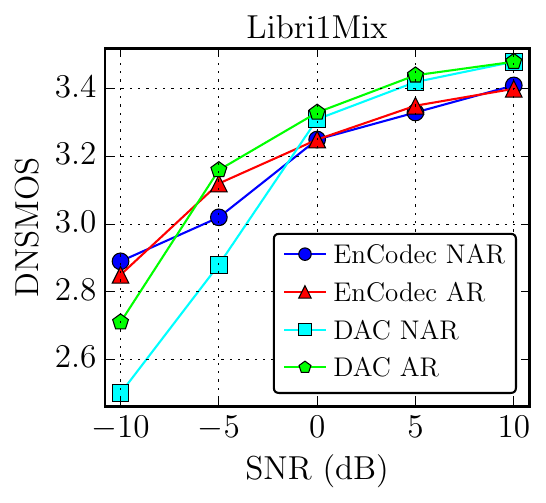}
    \end{subfloat}
    \end{center}
    \begin{center}
    \begin{subfloat}{}
    \includegraphics[width=0.20\textwidth]{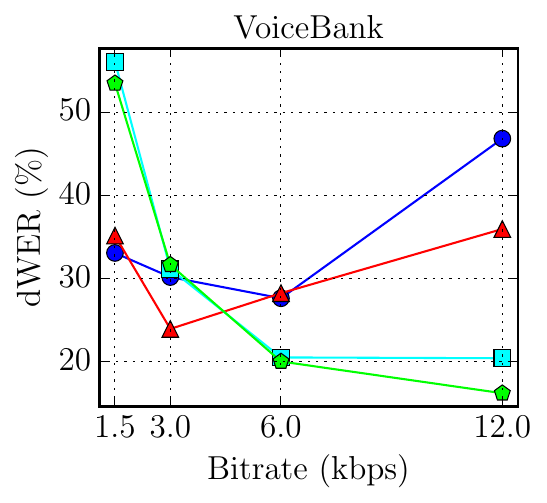}
    \end{subfloat}
    \begin{subfloat}{}
    \includegraphics[width=0.20\textwidth]{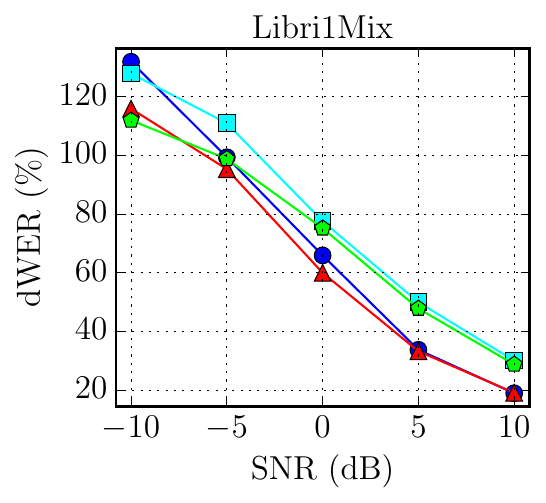}
    \end{subfloat}
    \end{center}
    \vspace{-.4cm}
\caption{\small {\textbf{(Left)}}: the effect of bitrate on VoiceBank. {\textbf{(Right)}}: the effect of noise strength on Libri1Mix.}
\label{fig:bitrate_noise}
\vspace{-0.65cm}
\end{figure}

\subsection{Bitrate and Noise Strength}
\label{subsec:bitrate}
\cref{fig:bitrate_noise} (Left) shows the effect of varying the codec's bitrate on VoiceBank. For DAC, we observe an improvement with respect to DNSMOS up to a bitrate of 6.0 kbps, after which it starts degrading, while the dWER keeps improving.
In contrast, for EnCodec, we observe saturation already at 3.0 kbps, beyond which both metrics quickly deteriorate.
This suggests that a complex trade-off exists between bitrate and capacity of the language model, which is highly dependent on the codec at hand.
In line with our previous findings, the SET architecture improves intelligibility. Notably, it performs particularly well in combination with DAC at high bitrates.

We also analyze the effect of noise strength in \cref{fig:bitrate_noise} (Right). A consistent degradation trend is observed for all codecs as the noise strength increases, with DAC's performance deteriorating more rapidly than EnCodec's for negative SNRs. Furthermore, it is noteworthy that our SET architecture alleviates the performance deterioration caused by strong noise, with significant improvements for DAC at SNRs of $-$5 and $-$10.

\begin{table}[t]
\setlength{\tabcolsep}{3pt}
\newcommand\TBstrut{\rule[-0.7ex]{0pt}{3.2ex}}
\centering
\vspace{-0.15cm}
\caption{\small Performance on VoiceBank using teacher forcing (TF), beam search (BS), and beam search after refinement (BSR).}
\vspace{-0.1cm}
\label{tab:exposure_bias}
\resizebox{0.47\textwidth}{!}{%
\begin{tabular}{l|c|c|c|c|c|c|c|c|c}
\hline
\multirow{2}{*}{\textbf{Method}} & \multicolumn{3}{c|}{\textbf{DNSMOS ($\uparrow$)}} & \multicolumn{3}{c|}{\textbf{CosSim ($\uparrow$)}} & \multicolumn{3}{c}{\textbf{dWER ($\downarrow$)}}  \\
\cline{2-10}
 & \textbf{TF} & \textbf{BS} & \textbf{BSR} & \textbf{TF} & \textbf{BS} & \textbf{BSR} & \textbf{TF} & \textbf{BS} & \textbf{BSR} \TBstrut \\ 
\hline  
\hline
\multirow{2}{*}{EnCodec AR} \TBstrut & \multirow{2}{*}{3.17} & \multirow{2}{*}{3.15} & \multirow{2}{*}{3.18} & \multirow{2}{*}{0.883} & \multirow{2}{*}{0.871} & \multirow{2}{*}{0.877} & \multirow{2}{*}{18.29} & \multirow{2}{*}{37.50} & \multirow{2}{*}{23.97} \\ 
 & & & & & & & & & \\ \hline
\multirow{2}{*}{\shortstack[l]{EnCodec AR \\ + Vocos}} \TBstrut & \multirow{2}{*}{\textbf{3.45}} & \multirow{2}{*}{\textbf{3.43}} & \multirow{2}{*}{\textbf{3.45}} & \multirow{2}{*}{\textbf{0.932}} & \multirow{2}{*}{\textbf{0.905}} & \multirow{2}{*}{\textbf{0.927}} & \multirow{2}{*}{\textbf{15.79}} & \multirow{2}{*}{{32.75}} & \multirow{2}{*}{\textbf{21.69}} \\
 & & & & & & & & & \\ \hline
\multirow{2}{*}{DAC AR} \TBstrut & \multirow{2}{*}{3.39} & \multirow{2}{*}{3.38} & \multirow{2}{*}{3.41} & \multirow{2}{*}{0.910} & \multirow{2}{*}{0.889} & \multirow{2}{*}{0.899} & \multirow{2}{*}{27.32} & \multirow{2}{*}{55.96} & \multirow{2}{*}{31.69} \\
 & & & & & & & & & \\ \hline
\multirow{2}{*}{\shortstack[l]{WavLM \\ discrete AR}} \TBstrut & \multirow{2}{*}{3.43} & \multirow{2}{*}{3.41} & \multirow{2}{*}{3.41} & \multirow{2}{*}{0.839} & \multirow{2}{*}{0.834} &  \multirow{2}{*}{0.833} & \multirow{2}{*}{20.81} & \multirow{2}{*}{\textbf{27.73}} & \multirow{2}{*}{22.77} \\
 & & & & & & & & & \\ \hline
\end{tabular}
}
\vspace{-0.5cm}
\end{table}

\subsection{Exposure Bias}
\label{subsec:exposure_bias}
Autoregressive language models are typically trained via teacher forcing, \ie they learn to predict the next token assuming correct predictions for all past tokens.
However, during inference, prediction errors can accumulate throughout the sequence, with catastrophic effect on performance. While experimental evidence indicates that such errors may not significantly impact language models in the text domains~\cite{he2021exposure}, our investigation shows that this phenomenon, known as \emph{exposure bias}, is non-negligible when working with audio tokens.
Indeed, as reported in \cref{tab:exposure_bias}, a considerable drop in all metrics is observed when using beam search (BS) compared to teacher forcing (TF). This gap is more substantial for acoustic tokenizers, since multiple tokens are predicted in parallel for each time step, thereby increasing the likelihood of at least one codebook prediction being incorrect.
However, the refinement step introduced in \cref{subsec:compared_methods} 
greatly reduces the impact of exposure bias, with beam search after refinement (BSR) narrowing the gap with teacher forcing.\looseness=-1

\section{Conclusions}
In this paper, we explored the use of acoustic tokens for speech enhancement and introduced a novel autoregressive transducer tailored for this task. Experiments on VoiceBank and Libri1Mix datasets show that acoustic tokens from EnCodec and DAC outperform semantic tokens from WavLM with respect to speaker identity preservation. Additionally, the proposed SET architecture improves speech intelligibility across various experimental settings. This architecture is also well-suited for streaming inference and hypothesis rescoring using external speech language models, opening up interesting directions for future research.
However, there is still a significant gap between discrete and continuous representations.
Next steps in the adoption of audio tokens involve developing more powerful discrete representations to help reduce this gap.
\looseness=-1

\bibliographystyle{IEEEtran}
\bibliography{refs}

\end{document}